\documentclass[aps,prc,twocolumn,superscriptaddress] {revtex4-1}
\usepackage{CJK}
\usepackage{graphicx}
\usepackage{hyperref}
\usepackage{amssymb}
\usepackage{amsmath}
\usepackage[normalem]{ulem}
\usepackage{url}
\usepackage{color}
\begin{document}

\begin{CJK*} {UTF8}{} 

\title{Impacts of momentum dependent interaction, symmetry energy and near-threshold $NN\to N\Delta$ cross sections on isospin sensitive flow and pion observables}

\author{Yangyang Liu}
\email{liuyangyang@ciae.ac.cn}
\affiliation{China Institute of Atomic Energy, Beijing 102413, China}

\author{Yingxun Zhang}
\email{zhyx@ciae.ac.cn}
\affiliation{China Institute of Atomic Energy, Beijing 102413, China}
\affiliation{Guangxi Key Laboratory of Nuclear Physics and Technology, Guangxi Normal University, Guilin, 541004, China}

\author{Junping Yang}
\affiliation{China Institute of Atomic Energy, Beijing 102413, China}

\author{Yongjia Wang}
\email{wangyongjia@zjhu.edu.cn}
\affiliation{School of Science, Huzhou University, Huzhou 313000, China}

\author{Qingfeng Li}
\email{liqf@zjhu.edu.cn}
\affiliation{School of Science, Huzhou University, Huzhou 313000, China}
\affiliation{Institute of Modern Physics, Chinese Academy of Sciences, Lanzhou 730000, China}

\author{Zhuxia Li}
\affiliation{China Institute of Atomic Energy, Beijing 102413, China}

\date{\today}

\begin{abstract}
Based on the ultra-relativistic quantum molecular dynamics (UrQMD) model, the impacts of momentum dependent interaction, symmetry energy and near-threshold $NN\to N\Delta$ cross sections on isospin sensitive collective flow and pion observables are investigated. Our results confirm that the elliptic flow of neutrons and charged particles, i.e. $v_2^n$ and $v_2^{ch}$, are sensitive to the strength of momentum dependence interaction and the elliptic flow ratio, i.e., $v_2^n/v_2^{ch}$, is sensitive to the stiffness of symmetry energy. For describing the pion multiplicity near the threshold energy, accurate $NN\to N\Delta$ cross sections are crucial. With the updated momentum dependent interaction and $NN\to N\Delta$ cross sections in UrQMD model, seven observables, such as directed flow and elliptic flow of neutrons and charged particles, the elliptic flow ratio of neutrons to charged particles, charged pion multiplicity and its ratio $\pi^-/\pi^+$, can be well described by the parameter sets with the slope of symmetry energy from 5 MeV to 70 MeV. To describe the constraints of symmetry energy at the densities probed by the collective flow and pion observables, the named characteristic density is investigated and used. Our analysis found that the flow characteristic density is around 1.2$\rho_0$ and pion characteristic density is around 1.5$\rho_0$, and 
we got the constrains of symmetry energy at characteristic densities are $S(1.2\rho_0)=34\pm 4$ MeV and $S(1.5\rho_0)=36\pm 8$ MeV. These results are consistent with previous analysis by using pion and flow observable with different transport models, and demonstrate a reasonable description of symmetry energy constraint should be presented at the characteristic density of isospin sensitive observables. 

\end{abstract}


\maketitle
\end{CJK*}

\section{Introduction}
\label{introduction}
The isospin asymmetric nuclear equation of state is crucial for understanding the isospin asymmetric objects, such as the structure of neutron-rich nuclei, mechanism of neutron-rich heavy ion collisions, the properties of neutron stars including neutron star mergers and core collapse supernovae\cite{BALi08,CJHorowitz2014JPG,Lattimer2004Science,Steiner_2010}. The symmetric part of isospin asymmetric equation of state has been well constrained by using the flow and Kaon condensation\cite{Danielewicz2002Sci}. However, the symmetry energy away from the normal density still have large uncertainty, and it leads that the constraint of symmetry energy becomes one of the important goal in nuclear physics\cite{Carlson2017,NuPECC17}.

The ultimate goal of symmetry energy constraint is to obtain the density dependence of symmetry energy over a wide range, and many efforts have been devoted to constrain the symmetry energy from subsaturation density to suprasaturation density. For probing the symmetry energy at suprasaturation density, the isospin sensitive observables in heavy ion collisions (HICs), such as the ratio of elliptic flow of neutrons to charged particles, hydrogen isotopes or protons ($v^n_2/v^{ch}_2$, $v^n_2/v^{H}_2$ or $v^n_2/v^{p}_2$)\cite{Russotto2011PLB,Cozma2013PRC,WangYJ2014PRC,Russotto2016PRC,Cozma2018EPJ} and the multiplicity ratio of charged pions (i.e., $M(\pi^-)/M(\pi^+)$ or named as $\pi^-/\pi^+$)\cite{BALi2002PRL,BALi2002NPA,ZGXiao2009PRL,ZQFeng2010PLB,WJXie2013PLB,JH2014PRC,Song15,Cozma2016PLB,YYLiu2021PRC,GCYong2021PRC,SpiRIT2021PRL}, were mainly used. By comparing the calculations to transverse-momentum-dependent or integrated FOPI/LAND and ASY-EOS elliptic flow data of nucleons and hydrogen isotopes, a moderately soft to linear symmetry energy is obtained with UrQMD\cite{Russotto2016PRC,Russotto2011PLB,WangYJ2014PRC} and T\"ubingen quantum molecular dynamics (T\"uQMD)  models\cite{Cozma2013PRC}. The lower limit of the slope of symmetry energy $L$ obtained with the flow ratio data is $L>60$ MeV\cite{Wang2020Fronp}, which overlaps with the upper limits of the constraints from nuclear structure and isospin diffusion, i.e., $L\approx 60\pm 20$ MeV\cite{BALi13,Oertel17,YXZhang2020PRC}. However, the constraints of symmetry energy from $\pi^-/\pi^+$ show strong model dependence\cite{ZGXiao2009PRL,ZQFeng2010PLB,WJXie2013PLB,JH2014PRC,Song15,Cozma2016PLB,YYLiu2021PRC,GJhang20}, and the extracted $L$ values ranges from 5 MeV to 144 MeV. It may be caused by the different treatments on the nucleonic potential, $\Delta$ potential, threshold effects, pion potential, Pauli blocking, in-medium cross sections and so on, and also by the different numerical technical for solving the transport equations.


To reduce the model dependence and enhance the reliability of the constraints of symmetry energy, especially at suprasaturation density, the transport model evaluations are required. The transport model evaluation project has made important progress on benchmarking the treatment of particle-particle collision\cite{YXZhang2018PRC,Akira2019PRC} and nucleonic mean field potential\cite{Maria2021PRC} in both Boltzmann-Uehling-Uhlenbeck (BUU) type and Quantum molecular dynamics (QMD) type models. For simulating the collisions or decay of resonance particles, the time-step-free method is suggested\cite{YXZhang2018PRC,Akira2019PRC} since this method automatically determine whether the resonance will collide or decay according to their collision time or decay time. In the UrQMD model, the time-step-free method is adopted in the collision part\cite{YXZhang2018PRC,Akira2019PRC}, and the nucleonic potential is also involved for extending its applications in low-intermediate energy HICs\cite{WangYJ2014PRC,YYLiu2021PRC}. This model has been successfully used to study the HICs from low-intermediate energy to high energies\cite{Bass1998PPNP,Bleicher1999JPG,WangYJ2014PRC,YYLiu2021PRC,Wang2020Fronp}. Another method to reduce the model uncertainties is simultaneously describing the observables data (or named doing combination analysis), such as isospin sensitive collective flow and pion observables. For the combination analysis on the isospin sensitive nucleonic and pion observables, there were few works to simultaneously investigate them except the T\"uQMD model\cite{Cozma2016PLB} as far as we know. Thus, it will be interesting to do combination analysis on nucleonic and pion observables back-to-back by the UrQMD model for increasing the reliability of the constraints of symmetry energy in the community.


In previous analysis on the neutrons to protons or to hydrogen isotopes elliptic flow ratios\cite{WangYJ2014PRC} or $\pi^-/\pi^+$ ratios\cite{YYLiu2021PRC} by UrQMD model, the momentum dependent interaction (MDI) form, i.e., $t_4\ln^2(1+t_5(\textbf{p}_1-\textbf{p}_2)^2)\delta(\textbf{r}_1-\textbf{r}_2)$, was used.  This form was extracted from the Arnold's optical potential data~\cite{Arnold1982PRC,Hartnack1994prc}. In 1990's, the real part of the global Dirac optical potential (Schr\"odinger equivalent potential) was published by Hama et al.~\cite{Hama1990PRC}, in which angular distribution and polarization quantities in proton-nucleus elastic scattering
were analyzed in the range of 10 MeV to 1 GeV. The Hama's data generated Lorentzian-type momentum-dependent interaction~\cite{Hartnack1994prc}, which give a stronger momentum dependent potential than the Arnold's form at high momentum, have been used in many version of transport 
models\cite{Isse2005PRC,LWChen2014EPJA,Nara2020PRC,Cozma2021EPJA,FZhang2022PRC,QFLi2009JPG} for studying high energy HICs. 
In another,  the cross sections of $NN\to N\Delta$ channel, i.e.,$\sigma_{NN \to N \Delta}$, used in UrQMD model are obtained by fitting CERN data \cite{Baldini1988}, and the fitting formula underestimate $\sigma_{NN\to N\Delta}$ near the threshold energy which will be shown in Figure~\ref{NNND-cross-section}. Thus, the refinements of MDI and formula of $NN\to N\Delta$ cross section $\sigma_{NN\to N\Delta}$ near the threshold are necessary for simultaneously describing the flow and pion observables. 

In this work, we will address these issues with the UrQMD model and investigate their influence on nucleonic flow and pion observables. Further, the constraints of symmetry energy at suprasaturation density are discussed with the updated version of UrQMD model. The paper is organized as follows: in Sect.\ref{model}, we briefly introduce the nucleonic potential, momentum dependent interaction and refined cross sections of $NN\to N\Delta$ channel. In Set.\ref{results}, the impacts of momentum dependent interaction, symmetry energy and refined $NN\to N\Delta$ cross sections on flow and pion observables are presented and discussed. By comparing the calculations with the ASY-EOS flow data and FOPI pion data, the constraints of symmetry energy at characteristic density are discussed. Sec.\ref{summary} is the summary of this work.

\section{UrQMD model}
\label{model}
The version of UrQMD model we used is the same as that in Ref.\cite{YYLiu2021PRC}, in which the cross sections of $N\Delta \to NN$ channel are replaced with a more delicate form by considering the $\Delta$-mass dependence of the M-matrix in the calculation of $N\Delta \to NN$ cross section\cite{YCui2019CPC}. This version has been successfully used to describe the FOPI experimental data of multiplicity and ratio of charged pion\cite{YYLiu2021PRC}, but did not use to simultaneously describe the pion and flow observables.

Since we focus on the effects of different forms of MDI, symmetry energy, and cross sections of $NN\to N\Delta$, we briefly introduce them in the following. The nucleonic potential energy $U$ is calculated from the potential energy density, i.e., $U=\int u d^3 r$. The $u$ reads as
\begin{eqnarray}
u & =& \frac{\alpha}{2}\frac{\rho^2}{\rho_0}+\frac{\beta}{\eta+1}\frac{\rho^{\eta+1}}{\rho_0^\eta}\\\nonumber
&& +\frac{g_{sur}}{2\rho_0}(\nabla \rho)^2+\frac{g_{sur,iso}}{\rho_0}[\nabla(\rho_n-\rho_p)]^2\\\nonumber
&&+u_{md}+u_{sym}.
\end{eqnarray}
The parameters $\alpha$, $\beta$, and $\eta$ are related to the two, three-body interaction term. The third and fourth terms are isospin independent and isospin dependent surface term, respectively. The $u_{md}$ is from the MDI term, and we will adopt two forms in this work. The $u_{sym}$ is the symmetry energy term. 

\begin{figure}[htbp]
\centering
\includegraphics[angle=0,scale=0.3]{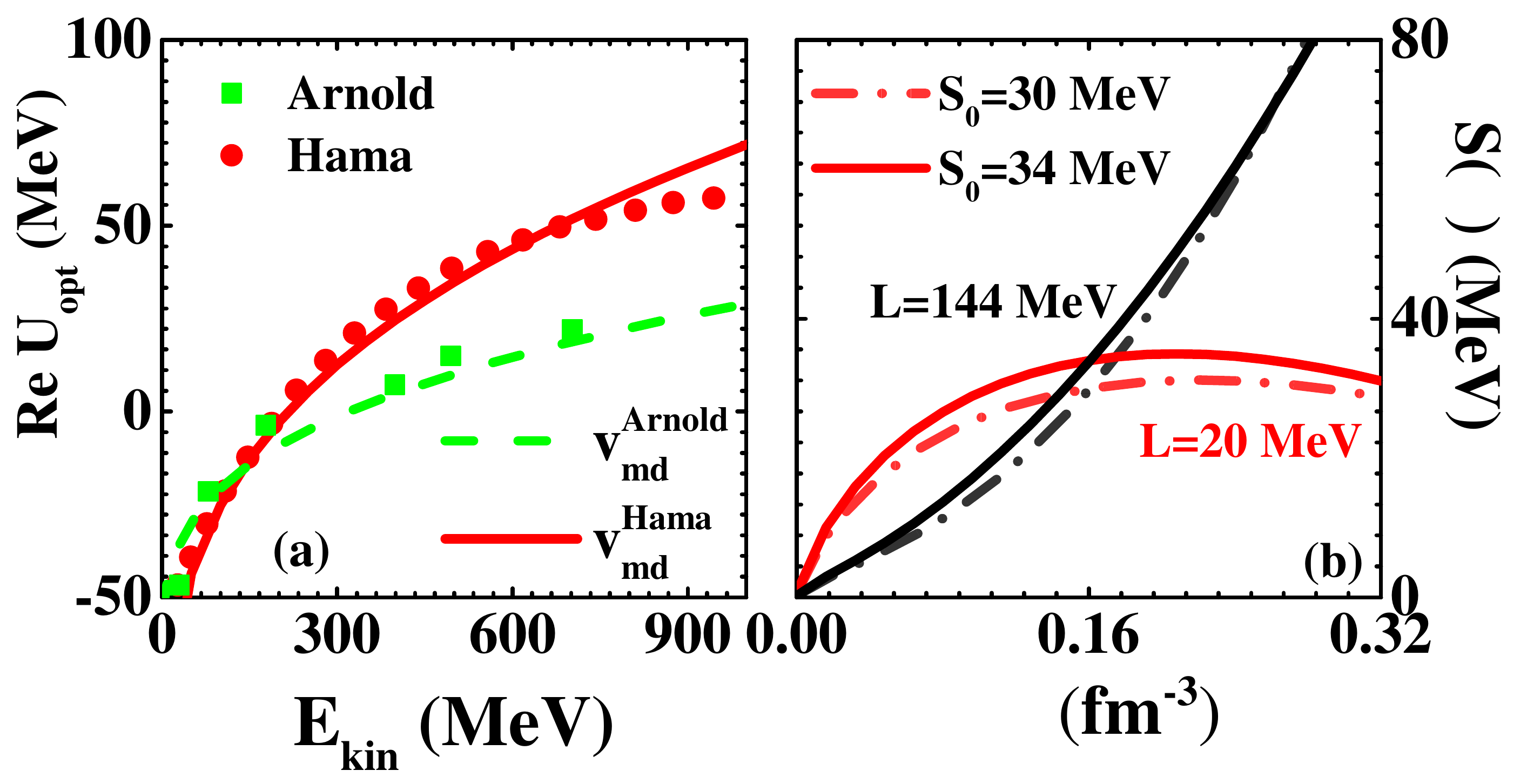}
\setlength{\abovecaptionskip}{0pt}
\vspace{2em}
\caption{(a) Real part of the optical potential $V_{md}$ and momentum dependent interaction $v_{md}$. The symbols are the optical potential data obtained from Arnold et al.~\cite{Arnold1982PRC} and Hama et al.~\cite{Hama1990PRC}. Lines are the $v_{md}^{Arnold}$ and $v_{md}^{Hama}$ obtained through Eq.(\ref{Vmdvmd}). (b) density dependence of the symmetry energy with different $S_0$ and $L$ values.}
\setlength{\belowcaptionskip}{0pt}
\label{Fig1Umd}
\end{figure}

The energy density associated with the MDI, i.e., $u_{md}$, is calculated according to the following relationship,
\begin{equation}
\begin{aligned}
u_{m d}=&\sum_{ij} \int d^{3} p_{1} d^{3} p_{2} f_{i}\left(\vec{r},\vec{p}_{1}\right) f_{j}\left(\vec{r},\vec{p}_{2}\right)v_{md}(\Delta p_{12}).
\end{aligned}
\end{equation}
The form of $v_{md}(\Delta p_{12})$ is assumed as,
\begin{equation}\label{Vmd-p}
	v_{md}(\Delta p_{12})= t_4\ln^2(1+t_5\Delta p_{12}^2)+C,
\end{equation}
where $\Delta p_{12}=|\textbf{p}_1-\textbf{p}_2|$, and the parameters $t_4$, $t_5$ and $C$ are obtained by fitting the data of the real part of optical potential. In details, we fit the data of real part of nucleon-nucleus optical potential $V_{md}(p)$ according to the following ansatz,
\begin{equation}\label{Vmdvmd}
   V_{md}(p_1)=\int_{p_2<p_F}v_{md}(p_1-p_2)d^3p_2/\int_{p_2<p_F}d^3p_2.
\end{equation}
This method is as the same as that in Ref.\cite{Hartnack1994prc}. Two sets of data of the real part of optical potential are used. One is from Arnold et al.~\cite{Arnold1982PRC} which were used in previous version of UrQMD\cite{WangYJ2014PRC,YYLiu2021PRC}. Another is from Hama et al.~\cite{Hama1990PRC}. They are presented as green squares and red circles in Fig.~\ref{Fig1Umd} (a), respectively. The lines are momentum dependence interaction $v_{md}(\Delta p_{12})$ at normal density obtained by fitting Arnold's or Hama's data by using Eq.(\ref{Vmd-p}) and Eq.(\ref{Vmdvmd}) within the kinetic energy $E_{kin}\approx$ 750 MeV. The values of $t_4$, $t_5$ and $C$ obtained from Arnold's data and Hama's data are listed in Table~\ref{tab:table1}. The momentum dependence of $v_{md}^{Hama}(\Delta p_{12})$ is stronger than that of $v_{md}^{Arnold}(\Delta p_{12})$, and the value of $v_{md}^{Hama}(\Delta p_{12})$ is higher than $v_{md}^{Arnold}(\Delta p_{12})$ at high momentum region. 

To keep the incompressibility of symmetric nuclear matter $K_0=231$ MeV for two different MDIs, the parameter $\alpha$, $\beta$, and $\eta$ are readjusted and the values of parameters and corresponding effective mass $m^*/m$ are listed in Table~\ref{tab:table1}.
\begin{table}[htbp] 
\caption{\label{tab:table1}%
Parameters used in the present work.  $t_4$, $C$, $\alpha$, $\beta$ and $K_0$ are in MeV. $t_5$ is in MeV$^{-2}$, $\eta$ and $m^*/m$ are dimensionless. The width of Gaussian wave packet is taken as 1.414 fm for Au+Au collision.}

\begin{tabular}{lcccccccc}
\hline
\hline
$Para.$ & $t_4$ & $t_5$ & $C$ &$\alpha$ & $\beta$ & $\eta$  & $K_0$ & $m^*/m$\\
\hline
$v_{md}^{Arnold}$ & 1.57 & 5$\times$10$^{-4}$ & -54  &-221 & 153 & 1.31   & 231 & 0.77 \\
$v_{md}^{Hama}$ & 3.058 & 5$\times$10$^{-4}$ & -86  &-335 & 253 & 1.16  & 231 & 0.635 \\
\hline
\hline
\end{tabular}
\end{table}

For the potential energy density of symmetry energy part, i.e., $u_{sym}$, we take two forms in the calculations. One is the Skyrme-type polynomial form ( (a) in Eq.~(\ref{srho-lyy})) and another is the density power law form ((b) in Eq.~(\ref{srho-lyy})). It reads,
\begin{eqnarray}
\label{srho-lyy}
 u_{sym}&=&S^{pot}_{sym}(\rho)\rho\delta^2\\\nonumber
 &=&\left\{
 \begin{array}{ll}
    ( A(\frac{\rho}{\rho_0})+B(\frac{\rho}{\rho_0})^{\gamma_s}+C(\frac{\rho}{\rho_0})^{5/3} )\rho\delta^{2}, & \mathbf{(a)}\\
    \frac{C_{s}}{2}(\frac{\rho}{\rho_{0}})^{\gamma_i}\rho\delta^2. & \mathbf{(b)}
  \end{array}
\right.
\end{eqnarray}
The symmetry energy coefficient is $S_0=S(\rho_0)$ and the slope of symmetry energy is $L=3\rho_0\partial S(\rho)/\partial \rho|_{\rho_0}$. Based on the values of $S_0$, $L$ and parameters in Table~\ref{tab:table1}, one can also obtain the parameters of Eq.(\ref{srho-lyy}) based on the relationship described in Ref.~\cite{YXZhang2020PRC,Zhang2020FOP}. In following calculations, we taken $S(\rho_0)=30-34$ MeV and $L=5-144$ MeV, as shown in Table.\ref{tab:table2}.

For $L<35$ MeV, we use the Skyrme polynomial form of $S^{pot}_{sym}(\rho)$ because the simple power law form of symmetry energy can not give reasonable values at subnormal density. Further, the $L < 5$ MeV sets are not adopted because the corresponding symmetry energy becomes negative at the densities above $2.7\rho_0$ and the EOS will not be favored by the properties of the neutron stars. Thus, the lower limit of $L$ in our calculations is 5 MeV. For $L>35$ MeV, we use the simple power law form of symmetry energy. As an example, we present the density dependence of symmetry energy in Fig.\ref{Fig1Umd} (b) for $L$ = 20, 144 MeV at $S_0$=30 and 34 MeV.  


\begin{table}[htbp]
\normalsize 
\caption{\label{tab:table2}%
Parameters of symmetry energy and effective mass used in the calculations.}
\begin{tabular}{lccc}
\hline
\hline
Para. Name & Values & Description  \\
\hline
$S_0$ & [30, 34]  & symmetry energy coefficient     \\
$L$ & [5,144]  & slope of symmetry energy    \\
$m^*/m$ & 0.635,0.77  &  isoscalar effective mass   \\
\hline
\hline
\end{tabular}
\end{table}

\begin{figure}[htbp]
\centering
\includegraphics[angle=0,scale=0.3]{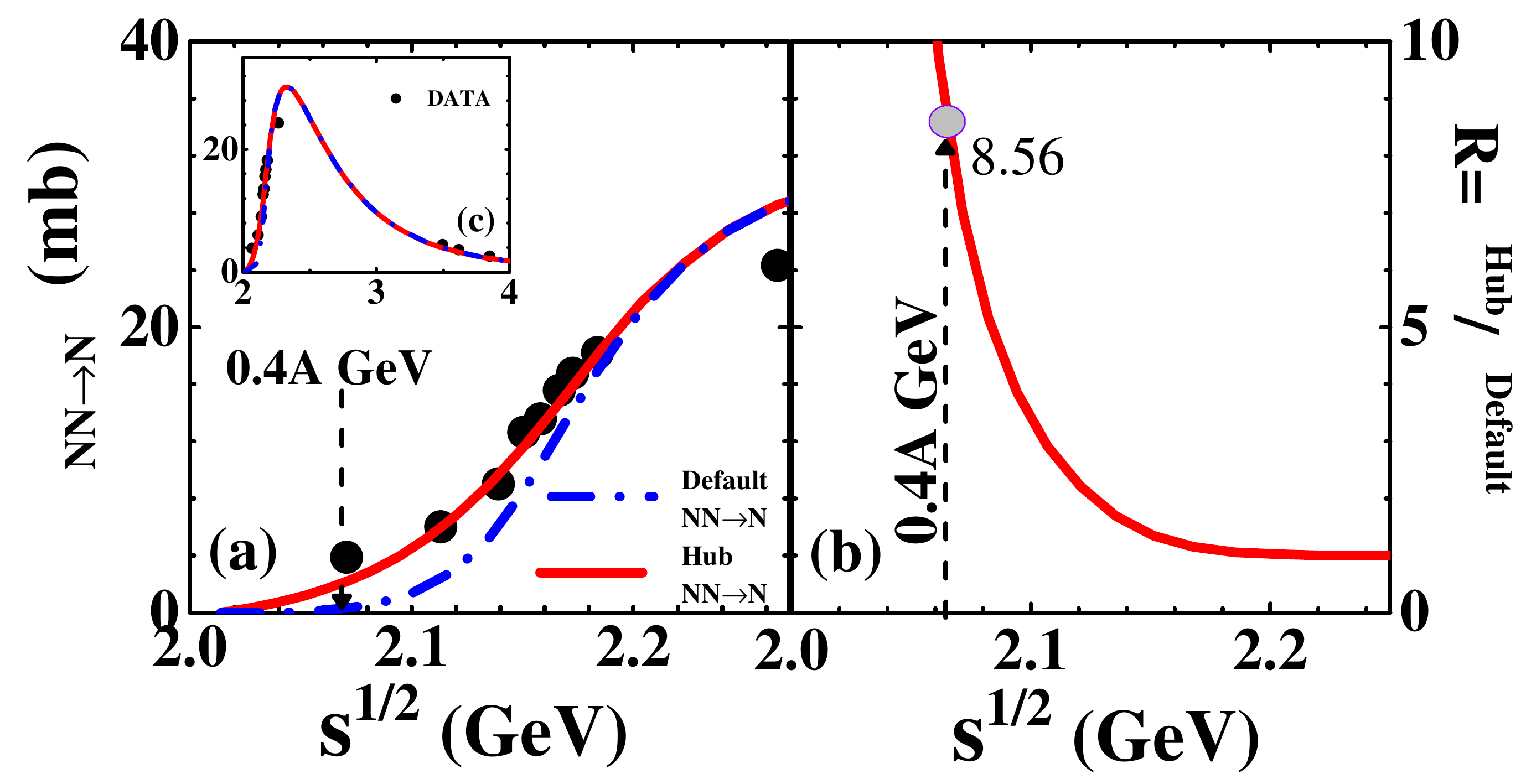}
\setlength{\abovecaptionskip}{0pt}
\vspace{2em}
\caption{(a) The cross section of {$NN\to N\Delta$} channel used in the default UrQMD model $\sigma_{NN\to N\Delta}^{Default}$ and obtained by refitting the experimental data with Hubbert function $\sigma_{NN\to N\Delta}^{Hub}$ near threshold energy. (b) The ratio of $\sigma_{NN\to N\Delta}^{ Hub}$ over $\sigma_{NN\to N\Delta}^{Default}$ as a function of $\sqrt s$.}
\setlength{\belowcaptionskip}{0pt}
\label{NNND-cross-section}
\end{figure}

In the collision term, the medium modified nucleon-nucleon elastic cross sections are used as the same as that in our previous works\cite{Wang2020Fronp}. For the $NN\to N\Delta$ cross sections, we found that the default formula used in UrQMD model in Ref.~\cite{Bass1998PPNP} underestimates the data~\cite{Baldini1988} near the threshold energy. The discrepancy is shown in Fig.\ref{NNND-cross-section} (a), where the blue line is the fitting formula in Ref.~\cite{Bass1998PPNP} and solid symbols are the data taken from Ref.~\cite{Baldini1988}. 

Thus, one can expect that we have to use an accurate form of $NN\to N\Delta$ cross section near the threshold energy for describing the pion production at 0.4A GeV. 
To refine the fitting of $NN\to N\Delta$ cross section near the threshold energy, we adopt a Hubbert function form to describe the $NN\to N\Delta$ cross sections at $\sqrt{s}<2.21$ GeV. That is,
\begin{eqnarray}
 \sigma_{N N \rightarrow N \Delta}(\sqrt{s})&=&A_1+\frac{4  A_2 * e^{-(\sqrt{s}-A_3)/ A_4}}{(1+e^{-(\sqrt{s}-A_3) / A_4})^2}, \\\nonumber
 &&\sqrt{s}<2.21 GeV.   
\end{eqnarray}
In which, $A_1$=-1.11 mb, $A_2$=26.30 mb, $A_3$=2.24 GeV, and $A_4$=0.05 GeV. We named it as $\sigma_{NN\to N\Delta}^{Hub}$ to distinguish the default form in Ref.\cite{Bass1998PPNP}. The fitting results are represented as the red line in Fig.\ref{NNND-cross-section} (a). Above 2.21 GeV, the original fitting function is used. 

As shown in Fig.\ref{NNND-cross-section} (a), the $\sigma_{NN\to N\Delta}^{Hub}$ is closer to the experimental data than the original formula. The right panels show that the ratio of $R=\sigma^{Hub}/\sigma^{Default}$, and one can see that the cross sections $\sigma_{NN\to N\Delta}^{Hub}$ are increased by a factor of 8.56 at the beam energy of 0.4A GeV. Consequently, one can expect a higher pion multiplicity with $\sigma_{NN\to N\Delta}^{Hub}$ than the one with $\sigma_{NN\to N\Delta}^{Default}$. The $N\Delta \to NN$ cross sections are obtained based on the detailed balance, in which a $\Delta$ mass dependent $N\Delta\to NN$ cross sections was also considered as in Refs.~\cite{YCui2019CPC,YYLiu2021PRC}.  


\section{Results and discussions}
\label{results}
The collective flow reflects the directional features of the transverse collective motion, and it can be quantified in terms of the moments of the azimuthal angle relative to the reaction plane, i.e., $v_n=\langle\cos(n\phi)\rangle$, $n=1, 2, 3, \cdots$. Among the $v_n$, the elliptic flow $v_2$ has been used to determine the MDI \cite{Danielewicz2000NPA}, and the ratio between $v_2$ of neutrons and protons, i.e., $v_2^n/v_2^p$, or ratio between $v_2$ of neutrons and charged particles, i.e., $v_2^n/v_2^{ch}$, are proposed to determine the symmetry energy at suprasaturation density \cite{WangYJ2014PRC,Russotto2016PRC,Cozma2018EPJ}. It is known that pions are mainly produced through $\Delta$ resonance decay in suprasaturation density region at early stage; and the multiplicity ratio of charged pions, i.e., $\pi^-/\pi^+$, was also supposed as a probe to constrain the symmetry energy at suprasaturation density and widely studied\cite{BALi2002PRL,BALi2002NPA,ZGXiao2009PRL,ZQFeng2010PLB,WJXie2013PLB,Cozma2016PLB,YYLiu2021PRC}. In this work, we first investigate the nucleonic flow observables to determine the form of MDI and pion production to determine the form of $NN\to N\Delta$ cross sections near the threshold energy. Then, the symmetry energy at suprasaturation density will be extracted by comparing the UrQMD calculations of $v_2^n/v_2^{ch}$ to ASY-EOS data and comparing $\pi^-/\pi^+$ results to FOPI data.

\subsection{collective flow and pion observable}

In this work, we perform the calculations of Au+Au collision at 0.4A GeV witsubsectionh 200,000 events at each impact parameter. The final observables are obtained by integrating over $b$ from 0 ot $b_{max}$ with a certain weight. The weight of $b$ is reconstructed by the centrality selection used in the experiments where the $Z_{bound}$ or $Z_{rat}$ and the detected charge particle multiplicity or the ratio of total transverse to longitudinal kinetic energies in the center-of-mass (c.m.) system are used as in Refs.~\cite{Russotto2016PRC,FOPI2010}.
The corresponding impact parameter distributes in a wide range and the weight of $b$ is a Gaussian shape rather than a triangular shape~\cite{Russotto2016PRC}, which also have been discussed in Refs.\cite{Frankland2021PRC,Lili2019PRC,Lili2022arxiv}. 
The seven observables are investigated in the following analysis, as listed in Table.\ref{observable}. 

\begin{table}[htbp]
\caption{\label{observable}%
Seven experimental observables used in this work.}
\begin{tabular}{lcccc}
\hline
\hline
$obsevable$ & rapidity $y_0$ cut & $\theta_{lab}$  cut & $<b>$ \\
\hline
$v^n_1(p_t/A)$ & $-0.5-0.5$ & $37^ {\circ}-53^ {\circ}$ &5.69 fm \cite{Russotto2016PRC}\\
$v^{ch}_1(p_t/A)$ & $-0.5-0.5$ & $37^ {\circ}-53^ {\circ}$ & $5.69$ fm\cite{Russotto2016PRC}\\
$v^{n}_2(p_t/A)$ & $-0.5-0.5$ & $37^ {\circ}-53^ {\circ}$ & $5.69$ fm\cite{Russotto2016PRC} \\
$v^{ch}_2(p_t/A)$ & $-0.5-0.5$ & $37^ {\circ}-53^ {\circ}$ & $5.69$ fm\cite{Russotto2016PRC}\\
$v_2^n/v_2^{ch}(p_t/A)$ & $-0.5-0.5$ & $37^ {\circ}-53^ {\circ}$ &  $5.69$ fm \cite{Russotto2016PRC}\\
$M(\pi)$ & $-$ & $-$ &  $<$2\footnote{We did not put the average b value here since experimental paper only provides $b/b_{max}<0.15$, which is obtained by estimating the impact parameter b from the measured differential cross sections for the
ERAT under a geometrical sharp-cut approximation. \label{1}}\cite{FOPI2010} \\
$\pi^-/\pi^+$ & $-$ & $-$ & $<$2\textsuperscript{\ref {1}}\cite{FOPI2010}\\
\hline
\hline
\end{tabular}
\end{table}

\begin{figure}[htbp]
\centering
\includegraphics[angle=0,scale=0.35]{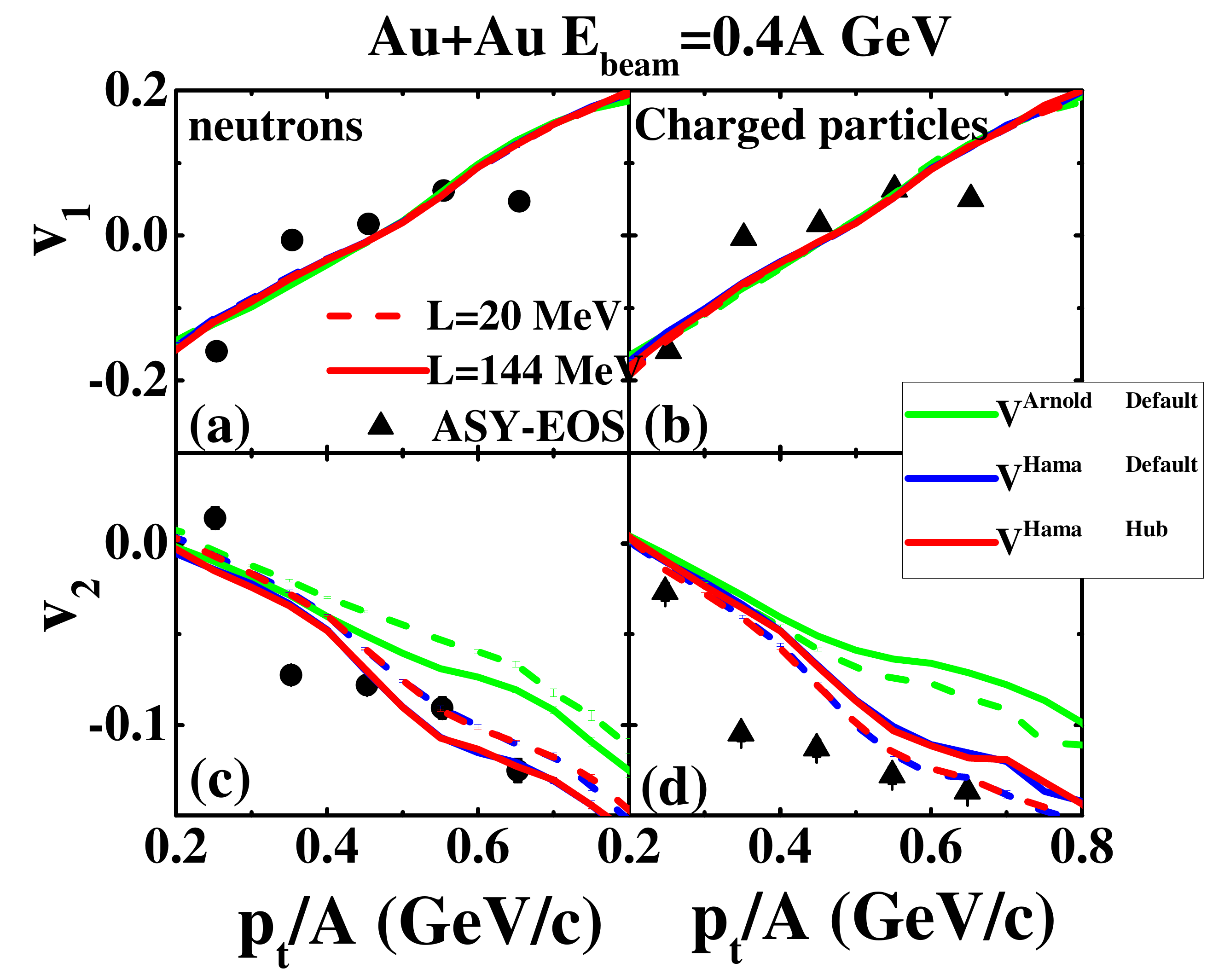}
\setlength{\abovecaptionskip}{0pt}
\vspace{2em}
\caption{ Panel (a) $v_1(p_t/A)$ for neutrons; (b) $v_1(p_t/A)$ for charged particles; (c) $v_2(p_t/A)$ for neutrons, and (d) $v_2(p_t/A)$ for charged particles. The green lines are for $V_{md}^{Arnold}$ and $\sigma^{Default}_{NN\to N\Delta}$, blue lines for $V_{md}^{Hama}$ and $\sigma^{Default}_{NN\to N\Delta}$, and red lines for $V_{md}^{Hama}$ and $\sigma^{Hub}_{NN\to N\Delta}$. The dash and solid lines represent the results with $L=20$ MeV and $L=144$ MeV. The ASY-EOS data of collective flow for neutrons and charged particles are shown as circle and triangle symbols\cite{Russotto2016PRC}.}
\setlength{\belowcaptionskip}{0pt}
\label{Fig4collective-flow}
\end{figure}

Fig.\ref{Fig4collective-flow} (a) and (b) show directed flow as a  function of $p_t/A$ for neutrons $v_1^n(p_t/A)$ and  for charged particles $v_1^{ch}(p_t/A)$ at given rapidity region and angle cut. The symbols are the ASY-EOS data from Ref.\cite{Russotto2016PRC}. The lines represent the results of UrQMD calculations with different forms of MDI, symmetry energy and different $NN\to N\Delta$ cross sections. The green lines are the results with $v_{md}^{Arnold}$ and $\sigma^{Default}_{NN\to N\Delta}$, blue lines are the results with $v_{md}^{Hama}$ and $\sigma^{Default}_{NN\to N\Delta}$. By comparing the green and blue lines, one can understand the effects of MDI. The red lines are the results for $v_{md}^{Hama}$ and $\sigma^{Hub}_{NN\to N\Delta}$. By comparing the blue lines and red lines, the effects of $\sigma_{NN\to N\Delta}$ can be understood. The dashed lines and solid lines represent the results with $L=20$ MeV and $L=144$ MeV at $S_0=32.5$ MeV, respectively. The calculations show that the $v_1^n(p_t/A)$ and $v_1^{ch}(p_t/A)$ increase from negative values to positive values with the increasing of $p_t/A$, and the sign of $v_1$ changes around $p_t/A\approx 0.5$ GeV/$c$. Furthermore, the calculations show that there is no sensitivities of $v_1$ to $L$, MDI and $\sigma_{NN\to N\Delta}$ at the selected rapidity region, due to the spectator matter blocking effect. In addition, the calculations with different combination of $L$, MDI and $\sigma_{NN\to N\Delta}$ falls in the data region.  

Fig.\ref{Fig4collective-flow} (c) and (d) show the elliptic flow for neutrons $v_2^n(p_t/A)$ and for charged particles $v_2^{ch}(p_t/A)$, with different $L$, MDI and $\sigma_{NN\to N\Delta}$. The symbols and lines have the same meaning as in panels (a) and (b). Both the $v_2^n$ and $v_2^{ch}$ have negative values and decrease with $p_t/A$ increasing, which means a preference for particle emission out of the reaction plane, towards $90^\circ$ and $270^\circ$. The important point is that both $v_2^n$ and $v_2^{ch}$ at high $p_t$ region are strongly sensitive to the strength of MDI and $L$, but hardly influenced by the forms of $\sigma_{NN\to N\Delta}$. The reason is that only 6\% of $NN$ collisions belong to $NN\to N\Delta$ collision in the present studied beam energy~\cite{YYLiu2021PRC}. The values of $v_2$ obtained with the $v_{md}^{Hama}$ are always lower than that with $v_{md}^{Arnold}$ due to the stronger momentum dependence of $v_{md}^{Hama}$ than that of $v_{md}^{Arnold}$. The calculations of $v_2^n$ and $v_2^{ch}$ with $v_{md}^{Hama}$ are more closed to the ASY-EOS experiment data than the one obtained with $v_{md}^{Arnold}$, which means that the $v_{md}^{Hama}$ is favored. Thus, the following analyzing on the symmetry energy effects are based on the MDI of $v_{md}^{Hama}$. 

In addition, both the $v_2^n$ and $v_2^{ch}$ exhibit some sensitivity to the stiffness of the symmetry energy. As shown in Fig.\ref{Fig4collective-flow} (c), the values of $v_2^n$ obtained with $L=144$ MeV (stiff) are lower than that with $L=20$ MeV (soft) case. The reason is that the stiff symmetry energy provides the stronger repulsive force on neutrons at suprasaturation density than that for soft symmetry energy cases. For charged particles, as shown in panel (d), $v_2^{ch}$ obtained with stiff symmetry energy case are higher than that with soft symmetry energy case. This is because the emitted charged particles are mainly composed of free protons, which feel stronger attractive interaction for stiff symmetry energy case than that for soft symmetry energy case at suprasaturation density.
However, $v_2^n$ or $v_2^{ch}$ cannot be used individually to constrain the symmetry energy, because both $v_2^n$ and $v_2^{ch}$ not only depend on the symmetry energy but also on the MDI and incompressibility. For example, the calculations with different incompressibility can lead to different results of the elliptic flow\cite{Wang2018PLB}. 



\begin{figure}[htbp]
	\centering
\includegraphics[angle=0,scale=0.32]{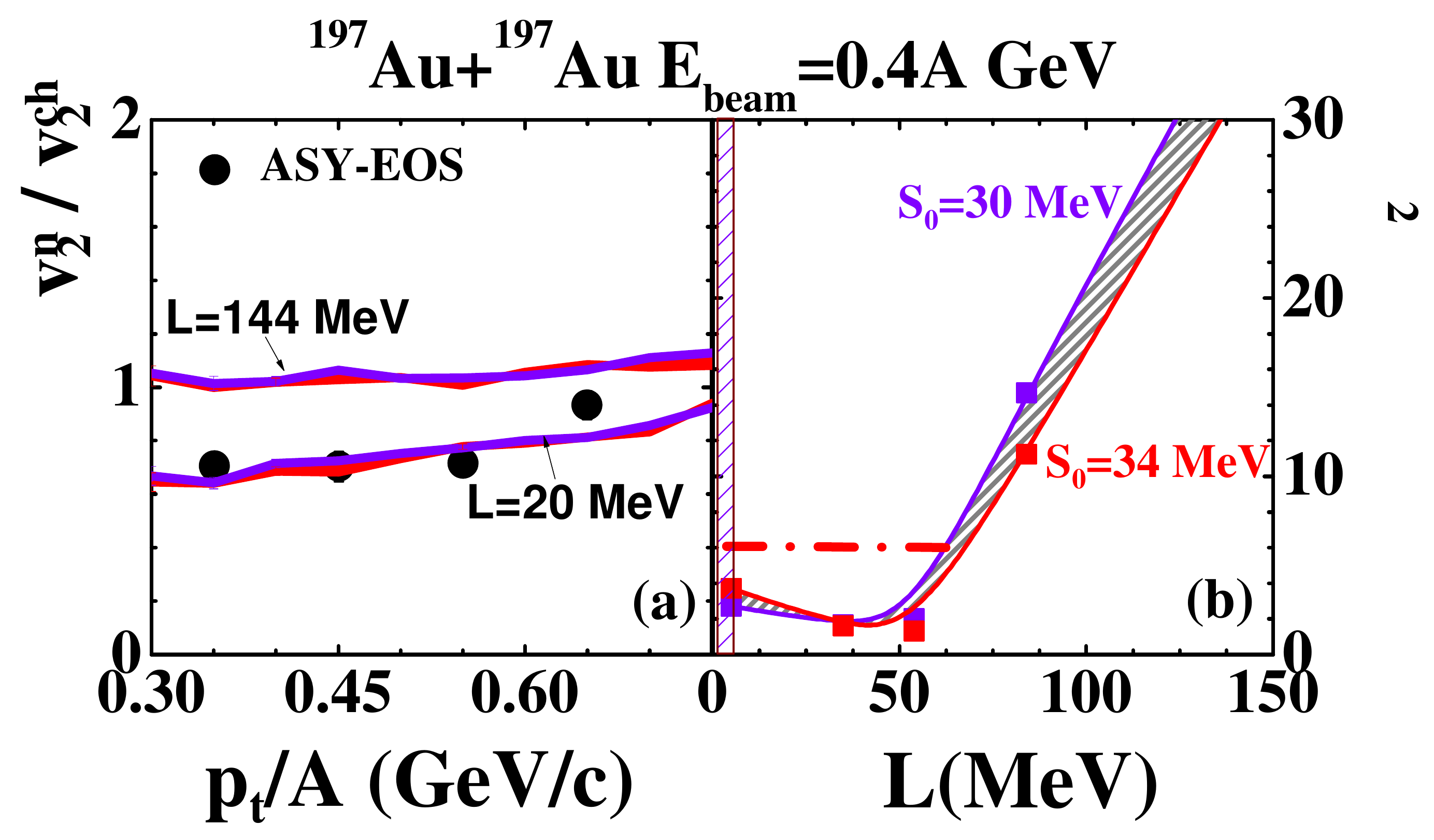}
	\setlength{\abovecaptionskip}{0pt}
	\vspace{2em}
	\caption{Panel (a) $v_2^n/v_2^{ch}$ as a function of $p_t$/A for $L=20$ MeV and 144 MeV at $S_0$=30 MeV (violet line) and 
 $S_0$=34 MeV (red line). The black symbols represent the ASY-EOS experimental data\cite{Russotto2016PRC}; (b) $\chi^2$ as a function of $L$ with different $S_0$. }
	\setlength{\belowcaptionskip}{0pt}
	\label{Fig3-v2Ratio}
\end{figure}

To isolate the contributions from the isocalar potential, $v_2^n/v_2^{ch}$ ratio was proposed to probe symmetry energy and several analysis have been performed by using the UrQMD model or T\"uQMD model\cite{Russotto2016PRC,Cozma2016PLB}. Fig.\ref{Fig3-v2Ratio} (a) shows the calculations for $v_2^n/v_2^{ch}$ as a function of $p_t$/A obtained with $v_{md}^{Hama}$. The symbols are the data points. The upper two lines are the calculations with $L=144$ MeV, and the lower two lines are for $L=20$ MeV. The violet lines are for $S_0$=30 MeV and red lines are for $S_0$=34 MeV. 
The calculations show that $v_2^n/v_2^{ch}$ is sensitive to $L$, especially at the low $p_t$ region in which the mean-field play more important role. The values of $v_2^n/v_2^{ch}$ obtained with stiff symmetry energy cases are larger than that with soft symmetry energy case. This behavior can be understood from Fig.\ref{Fig4collective-flow} (c) and (d). By comparing the calculations of $v_2^n/v_2^{ch}$ to ASY-EOS experimental data and doing a $\chi^2$ analysis, one can find the data favored parameter sets. In our work, the parameter sets are distinguished by the values of $S_0$ and $L$. Our conclusion is that the parameter sets with $L=5-70$ MeV and $S_0=30-34$ MeV can describe the data.

\begin{figure}[htbp]
	\centering
	\includegraphics[angle=0,scale=0.3]{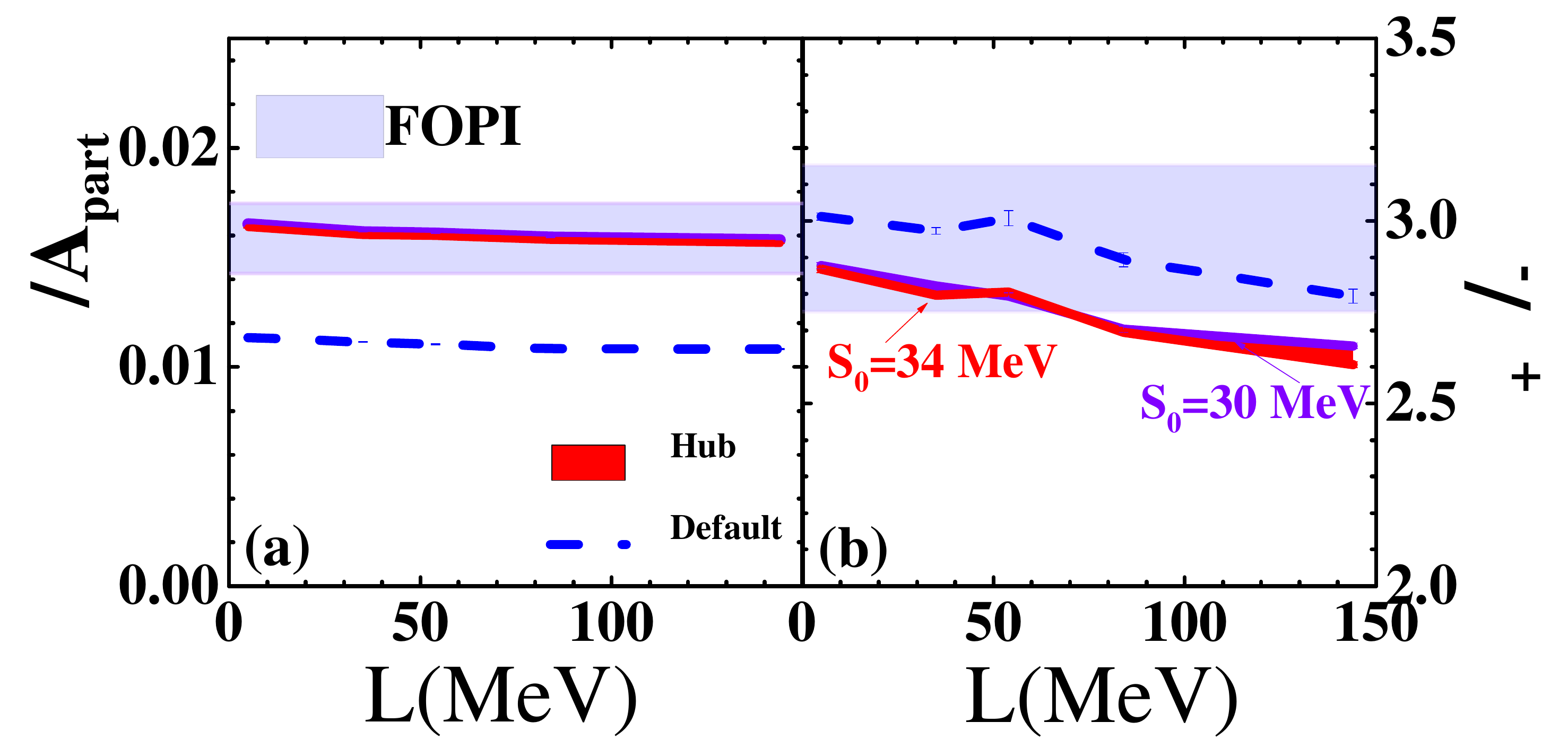}
	\setlength{\abovecaptionskip}{0pt}
	\vspace{2em}
	\caption{$M_{\pi}/A_{part}$ and $\pi^-/\pi^+$ as a function of $L$ with two forms of $\sigma_{NN\to N\Delta}$. The blue shaded region is the FOPI data\cite{FOPI2010}. The blue dashed lines represent the calculations obtained with $\sigma^{Default}_{NN\to N\Delta}$, and the  violet and red lines are the calculations with $\sigma^{Hub}_{NN\to N\Delta}$  for $S_0=30$ and 34 MeV.}
	\setlength{\belowcaptionskip}{0pt}
	\label{Fig5-MpionRatio}
\end{figure}

Fig.\ref{Fig5-MpionRatio} (a) shows the calculated $M_{\pi}/A_{part}$ as a function of $L$ with $v_{md}^{Hama}$, under different forms of $\sigma_{NN\to N\Delta}$. $A_{part}$ is the nucleon number of the participant, which is 90\% of the number of system. 
The blue lines represent the calculations obtained with $\sigma^{Default}_{NN\to N\Delta}$ in the UrQMD model.
By using the $\sigma^{Default}_{NN\to N\Delta}$, $M_{\pi}/A_{part}$ is underestimated by about 30\% relative to the data. This discrepancy can be understood from the underestimation of $NN\to N\Delta$ cross sections by using the default formula $\sigma_{NN\to N\Delta}^{Default}$ in UrQMD model, as shown in Fig.~\ref{NNND-cross-section}. The violet and red lines represent the results obtained with the $\sigma^{Hub}_{NN\to N\Delta}$ at $S_0$ varying from 30 to 34 MeV. The calculated results of $M_{\pi}/A_{part}$ fall into the data region since the $\sigma^{Hub}_{NN\to N\Delta}$ enhance the cross sections by a factor of 8.56 at 0.4A GeV relative to the default formula. But $M_{\pi}/A_{part}$ can not be used to probe $L$, since $M_{\pi}/A_{part}$ is insensitive to $L$ based on the calculations.


In Fig.\ref{Fig5-MpionRatio} (b), we present the calculated ratios $\pi^-/\pi^+$  as a function of $L$ with different forms of $\sigma_{NN\to N\Delta}$. Calculations show that $\pi^-/\pi^+$ is sensitive to $L$ for both forms of $\sigma_{NN\to N\Delta}$. Even the calculations with $\sigma_{NN\to N\Delta}^{Default}$ can reproduce the $\pi^-/\pi^+$ (blue line), one can not believe the conclusion since the pion multiplicity is underestimated relative to the data. For the calculations with $\sigma_{NN\to N\Delta}^{Hub}$, both the multiplicity of charged pion and its ratio $\pi^-/\pi^+$ can be reproduced. By comparing the calculations to the FOPI data, the parameter sets with $L=5-70$ MeV are also favored at $S_0=30-34$ MeV.

 \subsection{Characteristic density of nucleonic flow observable and symmetry energy constraints}
Before extracting the constraints of symmetry energy at suprasaturation density with collective flow and charged pion production, it is interesting to check the characteristic density probed by charged pion production and nucleonic flow observable. For pion observable, the characteristic density is obtained by averaging the compressed density with pion production rate and force acting on $\Delta$s in spatio-temporal domain in our previous work\cite{YYLiu2021PRC}, and the calculations show that the characteristic density of pion observable is around 1.5$\pm 0.5$ times normal density.

For the collective flow of neutrons and charged particles, the idea of calculating characteristic density is as same as pion characteristic density in our previous work\cite{YYLiu2021PRC}, but the weight is replaced by momentum change of nucleons. The momentum changes of nucleons during the time interval reflect the strength of the driven force for the collective motion of emitted particles, and can be used to understand the origins of $v_1$ and $v_2$. 
In the following calculations, two kinds of momentum change of nucleons are used. One is the momentum change in $x$-direction,
\begin{equation}\label{rho-ch-px}
\left\langle\rho_{\mathrm{c}, \text { flow }}\right\rangle_{\left|\Delta p_x\right|}=\frac{\int_{t_0}^{t_1} \Sigma_i\left|\Delta p_x^i(t)/\Delta t\right| \rho_c(t) d t}{\int_{t_0}^{t_1} \Sigma_i\left|\Delta p_x^i(t)/\Delta t\right| d t}
\end{equation}
and another is the momentum change in tranverse direction,
\begin{equation}\label{rho-ch-pt}
\left\langle\rho_{\mathrm{c}, \text { flow }}\right\rangle_{\left|\Delta p_t\right|}=\frac{\int_{t_0}^{t_1} \Sigma_i\left|\Delta p_t^i(t)/\Delta t\right| \rho_c(t) d t}{\int_{t_0}^{t_1} \Sigma_i\left|\Delta p_t^i(t)/\Delta t\right| d t}.
\end{equation}
The summation over $i$ runs over the nucleons belonging to the emitted nucleons and particles. More details, $|\Delta p^i_{x/t}(t)/\Delta t|=|(p^i_{x/t}(t)-p^i_{x/t}(t-\Delta t))/\Delta t|$, i.e., the momentum changes of nucleon during the time interval. $\rho_c(t)$ is obtained in a spherical region centered at c.m. of the system and with a radius of 3.35 fm. The region is used to represent the overlap region in semi-peripheral collisions of Au+Au. 

\begin{figure}[htbp]
\centering
\includegraphics[angle=0,scale=0.30]{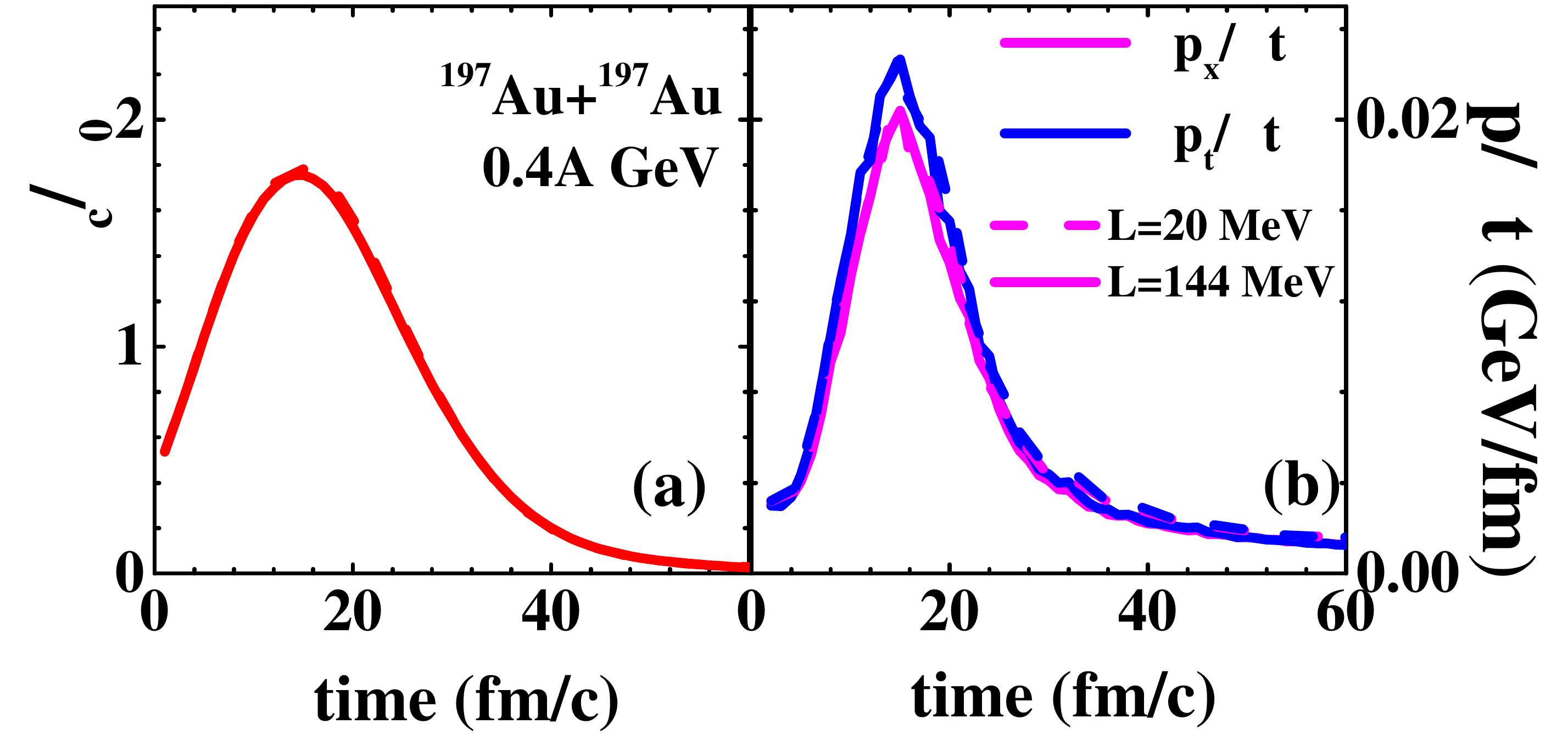}
\setlength{\abovecaptionskip}{0pt}
\vspace{2em}
\caption{(a) Time evolution of the averaged density in the center of
reaction system, (b) time evolution of momentum changes in $x$ direction $\Delta p_x/\Delta t$ and transverse direction $\Delta p_t/\Delta t$ .}
\setlength{\belowcaptionskip}{0pt}
\label{srho-flow}
\end{figure}

In Figure.\ref{srho-flow} (a), we plot the time evolution of the averaged central density $\rho_c(t)$ for a semi-peripheral collision of Au+Au. The averaged central density beyond normal density from 8 fm/$c$ to 28 fm/$c$ and reaches maximum values of 1.8$\rho_0$ at 16 fm/$c$ with the interactions we adopted. For convenience, we use $\Delta p/\Delta t$ to represent the momentum change per nucleon for nucleons and emitted particles, i.e.,
\begin{equation}
\frac{\Delta p}{\Delta t}=\frac{\Sigma_i\left|\Delta p^i(t)/\Delta t\right|}{N(t)}, 
\end{equation}
$N(t)$ is the total number of nucleons in the emitted nucleons and particles.
Panel (b) shows the average momentum changes of emitted particles $\frac{\Delta p}{\Delta t}$ as a function of time in $x$-direction and transverse direction. It illustrates that the drastic momentum changes of nucleons occur around 16 fm/$c$ when the participant region reaches the maximum density. It confirms that nucleonic flow observables mainly carry the EOS information at high density. Two forms of symmetry energy are tested, and they did not change the results dramatically.

\begin{figure}[htbp]
\centering
\includegraphics[angle=0,scale=0.55]{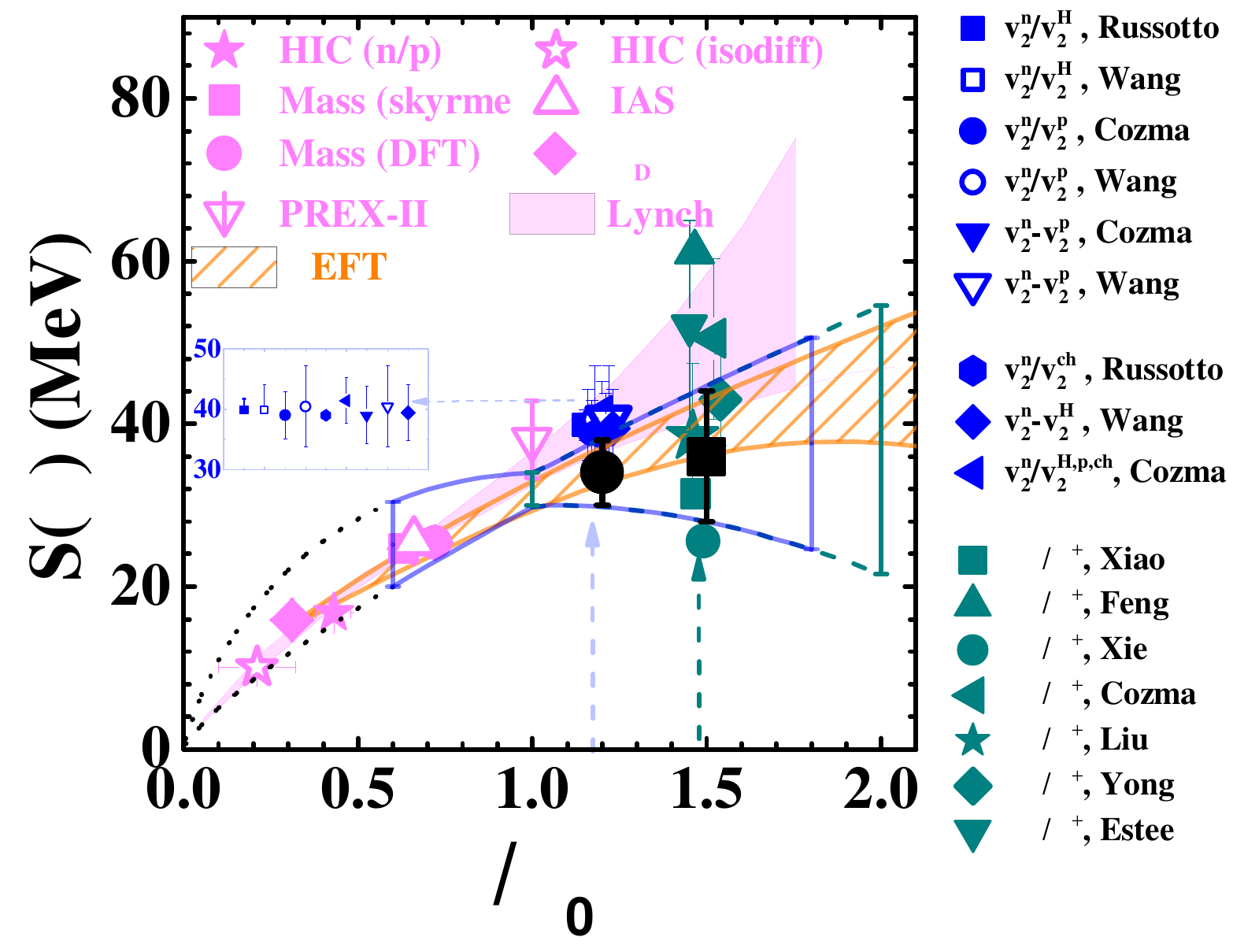}
\setlength{\abovecaptionskip}{0pt}
\vspace{2em}
\caption{ The constrains of the density dependence of symmetry energy at the collective flow characteristic density 1.2$\rho_0$ and the pion characteristic density 1.5$\rho_0$. }
\setlength{\belowcaptionskip}{0pt}
\label{Ssym-constrains}
\end{figure}

By using Eq.(\ref{rho-ch-px}) and Eq.(\ref{rho-ch-pt}), the characteristic density for the collective flow are obtained, and they are around $1.2\pm 0.6\rho_0$. It is consistent with the characteristic density obtained in the Ref.\cite{Ferve2016NPA} and Ref.\cite{gaobo}, but is smaller than the characteristic density obtained with pion observable. Thus, by comparing the calculations of $v_2^n/v_2^{ch}$ and $\pi^-/\pi^+$ to data, one can give the constraints of symmetry energy at two densities, i.e., 1.2 $\rho_0$ and 1.5 $\rho_0$. 
The values of them we got are $S(1.2\rho_0)=34\pm 4$ MeV and $S(1.5\rho_0)=36\pm 8$ MeV, and we present them as black symbols in Fig.\ref{Ssym-constrains}.  





The important point is that the constraints of $S(\rho)$ at flow characteristic density, i.e., at 1.2$\rho_0$, are consistent with the analysis of elliptic flow ratios or elliptic flow difference by UrQMD\cite{WangYJ2014PRC} or T\"{u}QMD calculations\cite{Cozma2013PRC,Cozma2018EPJ} which are presented by blue symbols. The constraints of $S(\rho)$ at pion characteristic density, i.e., at 1.5$\rho_0$, is consistent with our previous analysis\cite{YYLiu2021PRC} and constraints from the analysis of S$\pi$RIT by using dcQMD\cite{SpiRIT2021PRL} and isospin-dependent Boltzmann-Uehling-Uhlenbeck (IBUU)~\cite{GCYong2021PRC}, analysis of FOPI data by using T\"{u}QMD\cite{Cozma2016PLB}, IBUU~\cite{ZGXiao2009PRL} and isospin-dependent Boltzmann-Langevian (IBL)~\cite{WJXie2013PLB} within statistical uncertainties, except for the constraints obtained by Lanzhou quantum molecular dynamics (LQMD) model\cite{ZQFeng2010PLB}. 
Furthermore, if we extrapolate our constraints to subsaturation density, it also consists with the one at its characteristic density from the neutrons to protons yield ratios in HIC (n/p)\cite{Morfouace2019PLB}, isospin diffusion in HIC (isodiff)\cite{Tsang2008PRL}, mass calculated by the Skyrme\cite{Brown2013PRL} and density functional theory (DFT) theory\cite{Kortelainen2011PRC}, Isobaric analog state (IAS)\cite{Danielewicz2016NPA}, electric dipole polarization $\alpha_D$\cite{Zhang2015PRC},  at their sensitive density, which are decoded by Lynch and Betty in Ref.\cite{Lynch2022PLB}. Further, the extrapolated region is also consistent with the results from theoretical calculation with chiral effective field theory ($\chi$EFT)\cite{Drischler2020PRL}.

\section{Summary and outlook}
\label{summary}
In summary, we have investigated the influence of different momentum dependent interactions, symmetry energy and $NN\to N\Delta$ cross sections  on nucleonic observables and pion observables, such as $v_1^{n}$, $v_1^{ch}$, $v_2^{n}$, $v_2^{ch}$, $v_2^{n}/v_2^{ch}$, $M(\pi)$ and $\pi^-/\pi^+$, with UrQMD model for Au+Au at the beam energy of 0.4A GeV. Our results confirm that the elliptic flow of neutrons and charged particles, i.e. $v_2^n$ and $v_2^{ch}$, are sensitive to the momentum dependence potential. The ASY-EOS flow data favors the calculations with a strong momentum dependent interaction, i.e., $v_{md}^{Hama}$. However, the calculations with $v_{md}^{Hama}$ underestimate the pion multiplicity by about 30\% relative to FOPI data if the $\sigma^{Default}_{NN\to N\Delta}$ is adopted. Our calculations illustrate that the underestimation can be fixed by considering an accurate $NN \to N\Delta$ cross sections $\sigma^{Hub}_{NN\to N\Delta}$ in UrQMD model.

Further, the constraints on the symmetry energy at flow and pion characteristic densities are investigated with the updated UrQMD model. The characteristic density probed by flow is around 1.2$\rho_0$, which is smaller than the pion characteristic density 1.5$\rho_0$\cite{YYLiu2021PRC}. By simultaneously describing the data of $v_2^n/v_2^{ch}$ and $\pi^-/\pi^+$ with UrQMD calculations, the favored effective interaction parameter sets are obtained and we got the $S(1.2\rho_0)=34\pm 4$ MeV and $S(1.5\rho_0)=36\pm 8$ MeV. These results are consistent with previous analysis by using pion and flow observable with different transport models, and the consistency suggests that the reliable description of the constraints on symmetry energy should be presented at the characteristic density of isospin sensitive observables. 
By using more than one isospin sensitive observables which have different characteristic densities, the reliable of the extrapolation of symmetry energy at normal density can be enhanced. The extrapolated values of $L$ in this work are in $5-70$ MeV within $2\sigma$ uncertainty for $S_0=30-34$ MeV, which is below the analysis of PREX-II results with a specific class of relativistic energy density functional, but is consistent with the constrains from charged radius of $^{54}Ni$, from the combining astrophysical data with PREX-II and chiral effective field theory, and the S$\pi$RIT pion data for Sn+Sn at 0.27A GeV.

\section*{Acknowledgements}
The authors thank the discussions on the transport model and symmetry energy constraints at TMEP weekly meeting. This work was supported by the National Natural Science Foundation of China Nos.11875323, 12275359, 12205377, 11875125, U2032145, 11790320, 11790323, 11790325, and 11961141003, the National Key R\&D Program of China under Grant No. 2018 YFA0404404, the Continuous Basic Scientific Research Project (No. WDJC-2019-13), and the funding of China Institute of Atomic Energy (No. YZ222407001301), and the Leading Innovation Project of the CNNC under Grant No. LC192209000701, No. LC202309000201. We acknowledge support by the computing server C3S2 in Huzhou University.


\bibliography{ref}


\end{document}